\documentclass[10pt]{article}
\usepackage{amssymb,amsmath, psfrag, epsfig}
\def\o{\over}

\def\p{\partial}
\def\ov{\overline}
\def\h{\hat{X}}
\def\H{\hat{V}}

\def\b{\noindent}

\def\bs{\boldsymbol}
\def\c{\centerline}

\def\v{\vskip .15 in}
\def\ov{\overline}
\def\be{\begin{equation}}
\def\ee{\end{equation}} 

  \newcommand{\bi}{\bibitem}

\begin{document}

\title{{\bf Analysis of  inverse stochastic resonance and the
long-term firing of Hodgkin-Huxley neurons 
with Gaussian white noise}\\
{\small {\bf  Henry C. Tuckwell$^{1\dagger}$, J\"urgen Jost $^{2}$ }}\\   \
\  \\ 
{\small $^1$ Max Planck Institute for Mathematics in the Sciences\\
Inselstr. 22, 04103 Leipzig, Germany\\
$^{\dagger}$ {\it Corresponding author}: tuckwell@mis.mpg.de}}

\maketitle 

\newpage

\begin{abstract}
In previous articles we have investigated the firing properties
of the standard Hodgkin-Huxley (HH) systems of ordinary and partial
differential equations in response to input currents composed
of a drift (mean) and additive Gaussian white noise. For certain
values of the mean current, as the noise
amplitude increased from zero, the firing rate exhibited a minimum
and this phenomenon was called inverse stochastic resonance (ISR).
Here we analyse the underlying transitions from a stable
equilibrium point to the limit cycle and vice-versa.
Focusing on the case of a mean input current density
$\mu=6.8$ at which repetitive firing occurs and ISR had been found to be
 pronounced, some of 
the properties of the corresponding stable equilibrium point
are found. A linearized approximation around this point 
has oscillatory solutions from whose maxima spikes
tend to occur. A one dimensional diffusion is also
constructed for small noise based on the correlations between
the pairs of HH variables and the small magnitudes of the
fluctuations in two of them. 
 Properties of  the basin of attraction of the
limit cycle (spike) are investigated heuristically and also
the nature of distribution of spikes  at very small noise corresponding to
trajectories which do not ever enter the basin of attraction 
of the equilibrium point. Long term trials of duration 500000 ms are carried out
for values of the noise parameter $\sigma$ from
0 to 2.0, with results appearing in Section 3.
The graph of mean spike count versus $\sigma$ is 
divided into 4 regions $R_1,...,R_4,$ where $R_3$ contains
the minimum associated with ISR. In $R_1$ noise has 
practically no effect until a critical value of $\sigma = \sigma_{c_1}$ is reached.
At a larger critical value $\sigma = \sigma_{c_2}$, the probability of
transitions from the basin of attraction of the equilibrium point to
that of the limit cycle becomes greater than zero and the spike rate
thereafter increases with increasing $\sigma$. The quantitative scheme
underlying the ISR curve is outlined in terms of 
exit time random variables and illustrated diagrammatically.
In the final subsection 3.4, several statistical properties of the
main random variables
associated with long term spiking activity are given, including distributions
of exit times from the two relevant basins of attraction and the 
interspike interval.

\end{abstract}

\noindent {\it Short Title:} Hodgkin-Huxley   

\noindent \it Keywords and Phrases: Hodgkin-Huxley equations, noise
\rm 
\section{Introduction}
The Hodgkin-Huxley \cite{HOD}  systems of ordinary and partial differential equations,
based on the electrophysiology of the squid giant axon, are the cornerstone
of mathematical models of single neurons as well as several
types of cardiac cells.  Recent such studies include
those of Komendantov et al. \cite{KOM} for hypothathalamic 
magnocellular neuroendocrine cells, Saarinen et al.  \cite{SAR} for
cerebellar granule cells, 
Williams et al.  \cite{WIL} for ventricular myocytes,  
Kameneva et al.  \cite{KAM} for retinal ganglion cells  and  Drion et al.  \cite{DRI}
for dopaminergic neurons. Many of these
computational cell models contain 10 or more components as 
the important roles of many different ion channels have been discovered
since the appearance of the HH model. Analysis of such
higher-dimensional models is very complex as there may be 50 or
more parameters in distinction to the relatively few in the 4-component
Hodgkin-Huxley system. Since the latter does in fact embrace
some of the basic firing properties of neurons in general, there has
naturally been a large number of analyses and computational studies
of the HH systems. These include  both deterministic (for example,   \cite{HAS, BES, AIH, FUK, GUCK,CAL})
and stochastic 
 (for example,  \cite{HOR, BRO, TIE, AUS, OZE}) modeling.

In recent articles \cite{TUC4, TUC2, TUC3} we have explored the effects of 
both additive Gaussian white noise and conductance noise 
on repetitive firing  in
the Hodgkin-Huxley system. In the additive noise
case for the ordinary differential equation (ODE) model, 
when the mean input current density $\mu$ is 
not far above the threshold of 6.4 $\mu$A/cm$^2$ for repetitive
firing, the number of spikes in the first 500 or 1000 ms was
found to undergo a pronounced minimum (ISR) as the noise level
$\sigma$ increased from zero \cite{TUC4}. The minimum occurred
around $\sigma$ = 0.35. 
Guo \cite{GUO} has recently found similar results for the HH ODE system
with colored (Ornstein-Uhlenbeck process) noise. 
Similar results were found for
the partial differential equation (PDE) system \cite {TUC2, TUC3}  
where the spatial distribution of the noise was also an
important factor, which led to the disinction between the effects of 
noise on the instigation
and propagation of spikes.

\subsection{Model description}
In this article we restrict attention to the HH ODE system, which
corresponds to a  uniformly polarized or ``space-clamped'' neuron.
 The system of stochastic differential equations was given 
in our previous articles
but are repeated here for completeness and notation: 

\be
 dV = \frac{1}{C} [\mu + \overline{g}_Kn^4(V_K-V)  + \overline{g}_{Na}m^3h(V_{Na}-V)
       + g_L(V_L-V)]dt  + \sigma dW\ee
 and for the auxiliary variables
\be  dn= [\alpha_n(1-n) - \beta_nn]dt
 \ee
\be  dm= [\alpha_m(1-m) - \beta_mm]dt
 \ee
\be  dh= [\alpha_h(1-h) - \beta_hh]dt
 \ee
where $C$ is the membrane capacitance per unit area, $\mu$, which may depend on $t$, is the mean
           input current density, $\overline{g}_K$, $\overline{g}_{Na}$ and $g_L$ are the maximal (constant) potassium, 
           sodium and leak conductances per unit area with corresponding equilibrium potentials $V_K$, $V_{Na}$,
           and $V_l$, respectively. The noise enters as the derivative of a standard Wiener process $W$ and
           has amplitude $\sigma$. The auxiliary variables are $n(t)$, the potassium activation, $m(t)$, the sodium activation
           and $h(t)$, the sodium inactivation. The coefficients in the differential equations for the 
           auxiliary variables as functions of depolarization are
\be   \alpha_n(V)= {10-V \o 100[e^{(10-V)/10}-1]}
 \ee
\be   \beta_n(V)  = {1 \o 8} e^{-V/80}
 \ee
\be  \alpha_m(V) =  {25-V \o 10[e^{(25-V)/10}-1] }
 \ee
\be           \beta_m(V) = 4e^{-V/18}      
 \ee
\be   \alpha_h(V)={7 \o 100} e^{-V/20}
 \ee
\be  \beta_h(V)=  { 1\o e^{(30-V)/10} + 1}
 \ee

\section{Stable equilibrium point and limit cycle}
When $\mu$ is above the critical value $\mu_{c_1}$ for
 repetitive firing
(saddle-node bifurcation) 
and smaller than the value  $\mu_{c_2}$ at which there is a 
subcritical Hopf bifurcation, there are two
attractors consisting of a limit-cycle (action potential
trajectory) and a stable equilibrium point. 

\subsection{Stable equilibrium point $\boldsymbol x^*$}
Let us denote the random vector  $(V,n,m,h$) by  $\boldsymbol X =(X_1, X_2, X_3, X_4)$ and
rewrite the system of equations as 
\be dX_1=  F_1(\boldsymbol X) dt + \sigma dW \ee
\be dX_k= F_k(\boldsymbol X) dt \ee
where $k=2,3,4$. 
Because a mean input current density of $\mu$ =6.8 $\mu$A/cm$^2$
was found to exhibit a pronounced minimum in firing as
$\sigma$ increased,  we will focus
on this value of  $\mu$ unless stated otherwise.

Setting, with zero noise, 
\be \boldsymbol F (\boldsymbol x^*) = \bs 0  \ee
yields equilibrium points for the deterministic Hodgkin-Huxley system. 

The Jacobian matrix at the equilibrium point is defined as 
\be  J (\bs x^*) = \bigg\{ \frac { \p F_i }  {\p x_j   } \bigg\}_{\boldsymbol x^*} , i,j=1,...,4.  \ee

\centerline{\includegraphics[width=4 in]{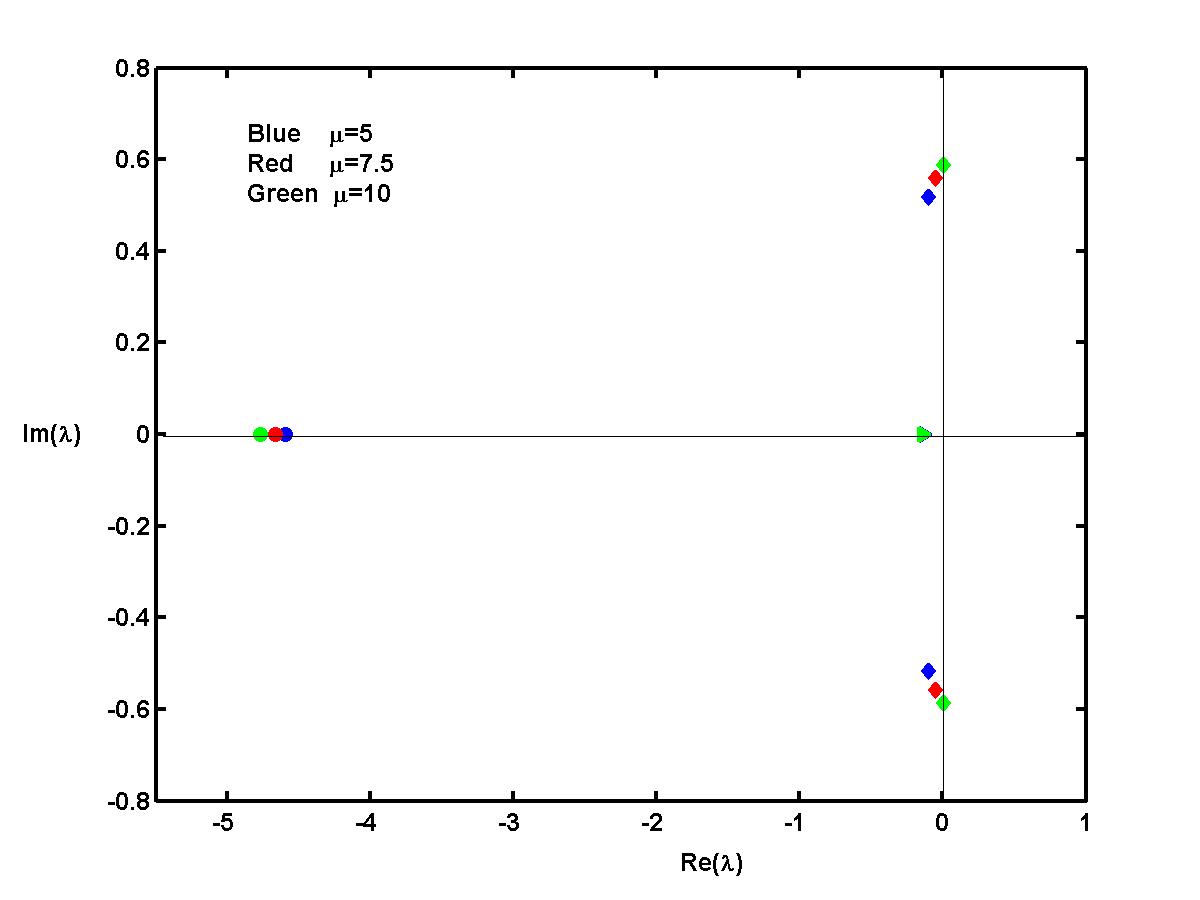}}
\begin{figure}[h]
\caption{Eigenvalues in the complex $\lambda$-plane for three
values of $\mu$. The values of the real eigenvalue near $\lambda= -0.13$ 
are indistinguishable in this diagram.  }
\label{}
\end{figure}
Figure 1 shows the eigenvalues in the complex lambda plane
for values of $\mu=5$, below the critical value  $\mu_{c_1}$,
$\mu=7.5$, between the critical values  $\mu_{c_1}$, and $\mu_{c_2}$,
and $\mu=10$ which is above  $\mu_{c_2}$. 
The nature of the critical point at the various values
of $\mu$ can be readily seen. 
For the chosen value of $\mu=6.8$, numerical evaluation 
gives an equilibrium point at
 \be \boldsymbol x^* =(4.0536, 0.38107, 0.084327, 0.45129) \ee
 at which the components of $F$ are
\be (-0.00000003084,   0.000024246,  -0.0000013509,  0.0000048941) \ee
Numerical evaluation of the eigenvalues of $J (\bs x^*) $ gives
$\lambda_1=-4.641$, the complex conjugate pair
$\lambda_{2,3}= -0.630 \pm 0.548i$ and $\lambda_4=-0.1323$.
Hence $\bs x^*$ is an asymptotically stable spiral point. 

\subsection{The linear approximation}

The system of stochastic ordinary differential equations for the
process  $\bs \h$
obtained by linearizing about $\bs x^*$ is
\be d \h_1 = \sum^4_{k=1} \frac { \p F_1 }  {\p x_k  } \h_k dt+ \sigma dW  \ee
\be d \h_j = \sum^4_{k=1} \frac { \p F_j }  {\p x_k  } \h_k dt, j=2,3,4,  \ee
where the partial derivatives are evaluated at $\bs x^*$. 
The Jacobian is found numerically to be 
\be 10^2 * \begin{bmatrix}
-0.010891&  -1.2764 &  1.2710
&  0.079467 \\
 0.000030551 &   -0.0019202 & 0
& 0 \\
 0.00032794 & 0& -0.034876 &
0 \\
-0.000044773 & 0 & 0 
&  -0.0012664
\end{bmatrix}
\ee

Note that the system (17)-(18) is linear and so $\bs \h$ is a Gaussian
process.  Using the theory in Rodriguez and
Tuckwell \cite{ROD}, in the absence of
an imposed spiking threshold,  the exact distribution of $\bs \h$  can
be found at any $t$.

Between spikes the linearized system (17)-(18) is expected to provide a reasonable
approximation to the fully nonlinear system (1)-(4). This is clearly demonstrated by
the two sets of sample paths shown in Figure 2. With input parameters 
$\mu=6.8$, $\sigma=0.6$, a time segment of length about 60 ms gave
the sample path for $V$ ($X_1$) shown in the top part of the Figure.
With the same path for the Wiener process (or the white noise),   
the sample path for the voltage $\hat{X}_1$ in the linearized system, with the same initial
values, is seen in the lower part of Figure 2,  to mimic closely that in the upper
part. However, there is one very striking difference betwen the two paths
as at about $t=59$ ms the voltage in the linear system attains a local
maximum and then decreases quite rapidly, continuing in an oscillatory fashion.
In contrast, at about the same $t$ and voltage values,  a spike arises in
the nonlinear system. However, these paths indicate that the time of spiking
in the nonlinear system can be well approximated as the time at which the
voltage in the linear sytem first attains a threshold value.  Such a first passage
time can be determined from the usual first exit time theory for diffusion
processes.

\centerline{\includegraphics[width=4 in]{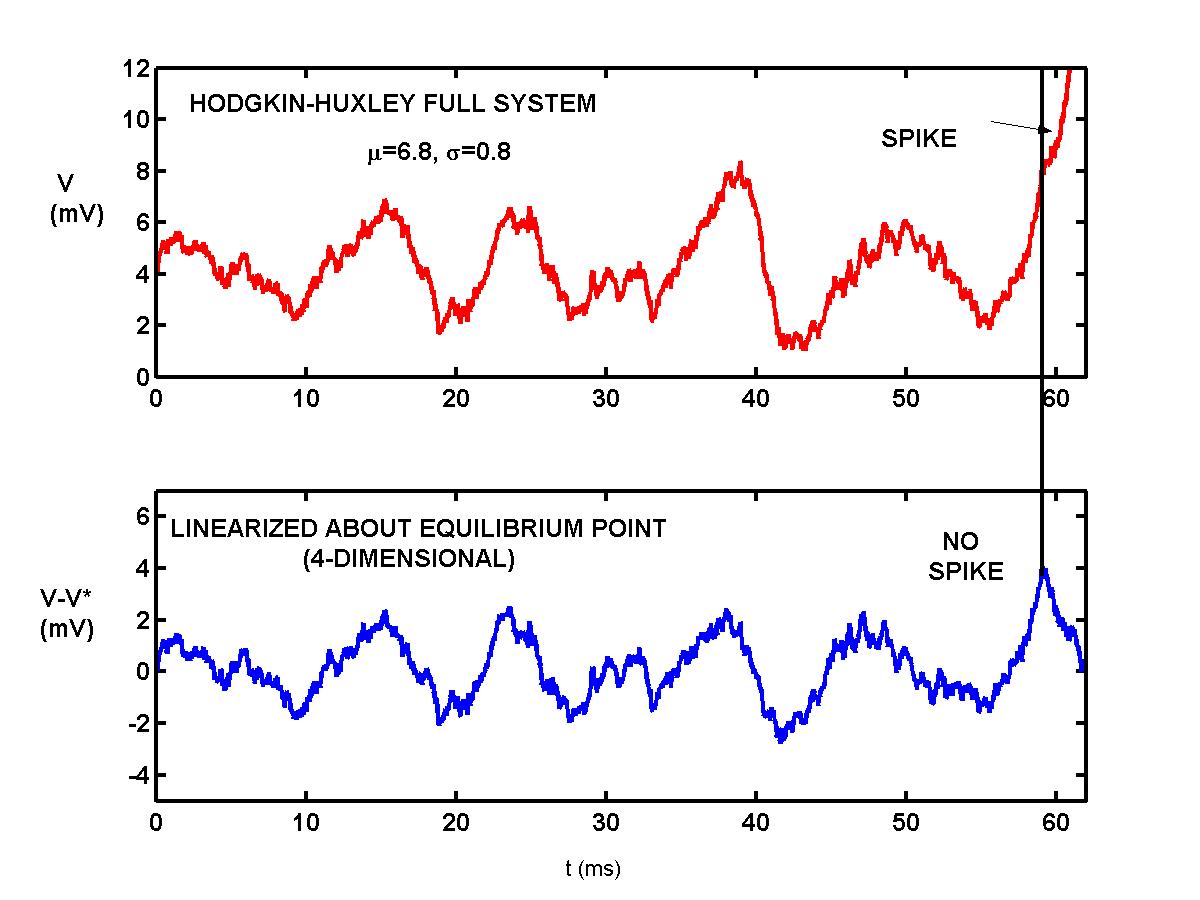}}
\begin{figure}[!ht]
\caption{In the top part, voltage  is plotted versus time 
for a sample path in the nonlinear full Hodgkin.Huxley system
with the parameters shown.  A spike forms near the end of
the record. In the lower record is shown the voltage path
for the system linearized about the stable equilibrium point.
The Wiener path is the same in both records to enable a
comparison to be made.}
\label{}
\end{figure}

\subsection{Numerical examples}
Without an imposed threshold condition, no spikes are possible in the linearized stochastic system 
described by (17) and (18) because there is only a stable equilibrium point
about which trajectories fluctuate.  This is of course in distinction to the
Hodgkin-Huxley system (with suitable input parameters) where trajectories
may, if the fluctuations are large enough, give rise to spikes around the
limit cycle. 
It is of interest to examine some statistical properties
of the original process $\boldsymbol X$ in nonspiking periods. 
An example of paths with $\mu=6.8$ and $\sigma=0.1$ over a 50 ms
nonspiking period is shown in Figure 2. The basic statistical properties 
of the components over a 500 ms time period are given in Table 1.
In Table 2 are given the corresponding correlation coefficients.

\begin{center}
\begin{table} [!h]
    \caption{Statistics of variables during non-spiking}
\c{period. $\mu=6.8$, $\sigma=0.1$}

\smallskip
\begin{center}
\begin{tabular}{|c|c|c|c|c|c|}

\hline
&Mean & Max  & Min  & St Dev & Coef Var\\
\hline
V& 4.049  & 4.446 &  3.658 &  0.1337 & 0.0330\\
n&   0.3811  & 0.3828 &   0.3795  & 6.61e-4 &  0.0017 \\
 m&   0.0843&   0.0874 &  0.0811 &  0.0012 & 0.0143\\
  h&  0.4515 & 0.4542 &  0.4488 &  0.001   & 0.0023\\

\hline
\end{tabular}
\end{center}
\end{table}
\end{center}

\begin{center}
\begin{table}
    \caption{Correlation coefficients during non-spiking}
\c{period. $\mu=6.8$, $\sigma=0.1$}

\smallskip
\begin{center}
\begin{tabular}{|c|c|c|c|c|}

\hline
&V & n  &  m  & h \\
\hline
V& 1.0000   & 0.3068 &   0.9562 &  -0.2137\\
n&   0.3068   & 1.0000 &   0.4649 &  -0.9894\\
 m&   0.9562 &   0.4649 &   1.0000 &  -0.3699 \\
  h&  -0.2137 & -0.9894 &  -0.3699 &   1.0000\\

\hline
\end{tabular}
\end{center}
\end{table}
\end{center}

\centerline{\includegraphics[width=4 in]{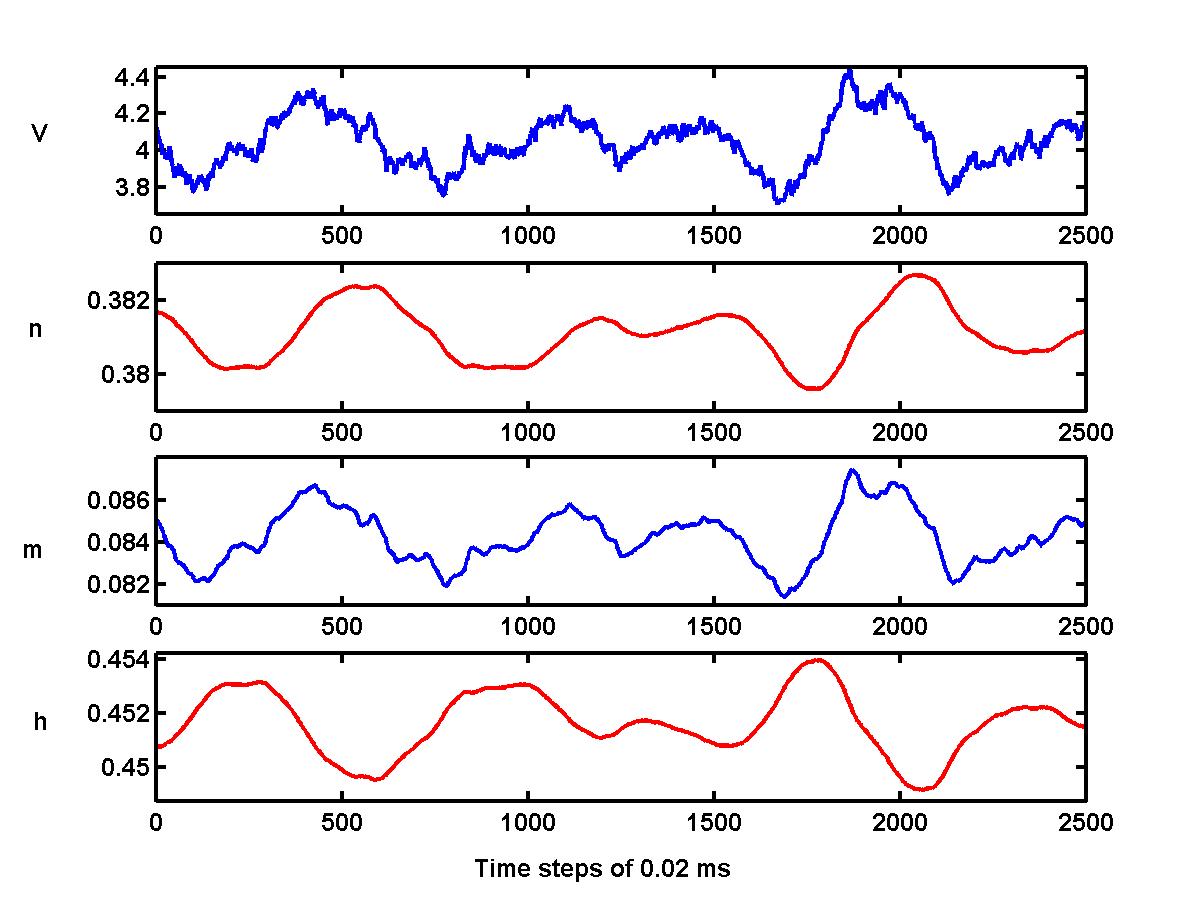}}
\begin{figure}[!hb]
\caption{Sample paths for the 4 components of $\boldsymbol X$ over a 
50 ms period during which there were no spikes. Input parameters
$\mu=6.8$ and $\sigma=0.1$.}
\label{}
\end{figure}

Tables 3 and 4 give the statistical properties during a 500 ms nonspiking
time period when the noise level is $\sigma=0.6$.

\begin{center}
\begin{table}[!ht]
    \caption{Statistics of variables during non-spiking}
\c{period. $\mu=6.8$, $\sigma=0.6$}

\smallskip
\begin{center}
\begin{tabular}{|c|c|c|c|c|c|}

\hline
&Mean & Max  & Min  & St Dev & Coef Var\\
\hline
V& 4.149  & 8.962 & -0.206 &  1.301 & 0.314\\
n&  0.383  & 0.419 &   0.368  & 0.0077 &  0.020 \\
 m&  0.0859 &   0.1393 &  0.0538 &  0.0126 & 0.147\\
  h&  0.4471 & 0.4668 &  0.3883 &  0.0122   & 0.0274\\

\hline
\end{tabular}
\end{center}
\end{table}
\end{center}

\begin{center}
\begin{table}
    \caption{Correlation coefficients, $\mu=6.8$, $\sigma=0.6$}
\smallskip
\begin{center}
\begin{tabular}{|c|c|c|c|c|}

\hline
&V & n  &  m  & h \\
\hline
V& 1.0000   & 0.3505 &   0.9690 &  -0.2408\\
n&   0.3505   & 1.0000 &   0.5284 &  -0.9854\\
 m&   0.9690 &   0.5284 &   1.0000 &  -0.4272 \\
  h&  -0.2408 & -0.9854 &  -0.4272 &   1.0000\\

\hline
\end{tabular}
\end{center}
\end{table}
\end{center}

There are two very noticeable features of the sample paths in the
nonspiking periods examined. Firstly, the oscillatory character
of the paths, which is traceable to the eigenvalues of the Jacobian
at the equilibrium point. Secondly, the strong positive correlation
between $V$ and $m$ and the strong negative correlation
between $n$ and $h$.

\centerline{\includegraphics[width=4in]{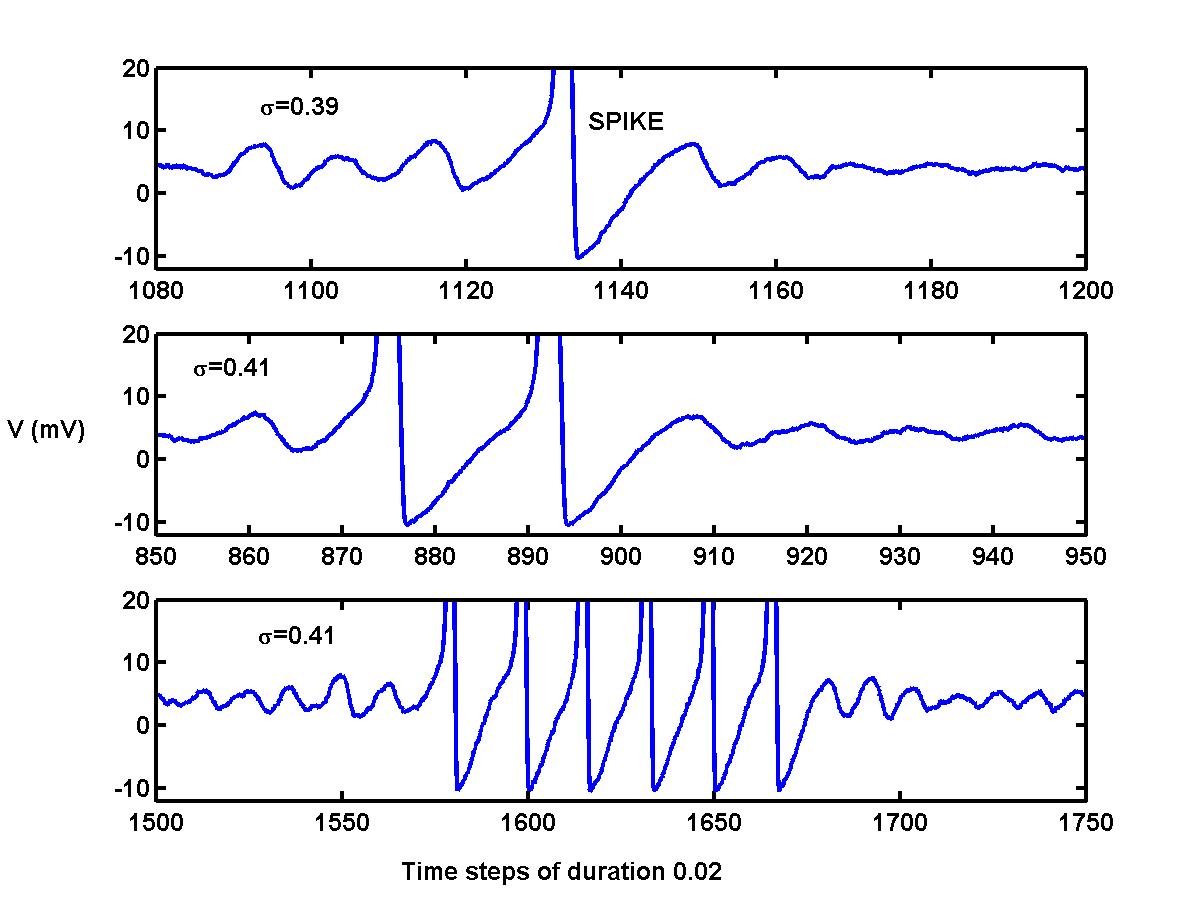}}
\begin{figure}[!hb]
\caption{Showing how single, double and multiple 
spikes arise from oscillations in the  nonlinear full Hodgkin-Huxley system with 
noise.  $\mu=6.8$}
\label{}
\end{figure}
 The oscillatory nature of the paths can
never be captured in the standard integrate and fire models
nor the leaky integrate and fire models \cite{TBOOK}. 
When the noise is sufficiently large to make for fairly frequent
spiking, the times of spiking must tend to arise at the maxima in the
oscillations of $V$. This is seen dramatically in Figure 3 
and in Figure 2 of  the previous subsection. 
\subsection{A one-dimensional diffusion}
Examination of the statistical properties of 
the process as given in Tables 1 and 2 shows that
for small noise the coefficients of variation of
$n$ and $h$ are an order of magnitude smaller than
those for $V$ and $m$ and that the correlation coefficient
of $V$ and $m$ is close to unity. These observations 
make it reasonable to consider a 1-dimensional approximation $\hat{V}$ to
the HH system in which a constant $\ov{n}  \approx n$ and a constant
$\ov{h}  \approx h$, with $m=kV$, where $k$ is another constant.
These approximations lead, on putting $C=1$ in (1), to the following 
stochastic differential equation for $\hat{V}$, 
\be d\H = (\mu + c_1 - c_2\H + c_3\H^3 - c_4\H^4)dt + \sigma dW, \ee
where
\be  c_1=  \overline{g}_K\ov{n}^4V_K + g_LV_L \ee
\be  c_2=  \overline{g}_K\ov{n}^4 + g_L  \ee
\be c_3=  \overline{g}_{Na}k^3\ov{h}V_{Na} \ee
\be  c_4=  \overline{g}_{Na}k^3\ov{h}.  \ee
Standard theory gives for such a diffusion that 
the mean exit time from a value $x\in (a,b)$ 
to outside this interval satisfies the ordinary differential equation
\be (\sigma^2/2)M''+  (\mu + c_1 - c_x + c_3x^3 - c_x^4)M'=-1, x\in (a,b), \ee
where primes denote differentiation, along with suitable boundary 
conditions. Preliminary investigations of the validity of this
approximation were made using values for 
$\mu=6.8$ and $\sigma=0.1$ (see Tables 1 and 2)
 but a detailed study will be reported 
in a subsequent article.

\subsection{The limit cycle and its basin of attraction}
With constant input current density in the interval
  $[\mu_{c_1}, \mu_{c_2})$, repetitive spiking may occur
with a fixed period. For $\mu=6.8$ and no noise
 the period is about 17.65 ms. 
 The limit cycle is in 4-space but we here show in the upper
part of Figure 2 the projection of the 
limit cycle obtained by plotting $n$ versus $V$.
In the lower part of the figure is shown the position
of the stable equilibrium point, here designated R, 
for the same value of $\mu$. It can be seen that the 
limit cycle approaches quite close to R.  
In a previous article in which the moment method
was used to explore the effects of noise
on HH spiking  (Tuckwell and Jost, 2009), we heuristically
estimated the basin of attraction of the stable equilibrium
point. Part of the basin of attraction of the limit cycle 
can be numerically estimated by taking the union of
all stochastic paths which do not collapse to the stable equilibrium
point.  As explained later, there are ranges of values of
$\sigma$ where the stochastic paths have this property 
and Figure 6 shows a sample of such paths for various
such $\sigma$. Here paths were taken over 1000 msec
involving about 50 spikes for values of $\sigma$ ranging
from 0.05 to 0.25.  This picture gives an idea of how
far off the deterministic limit cycle a path may wander without
entering the basin of attraction of the stable equilibrium point. 
The variability of paths is relatively small.

\centerline{\includegraphics[width=4 in]{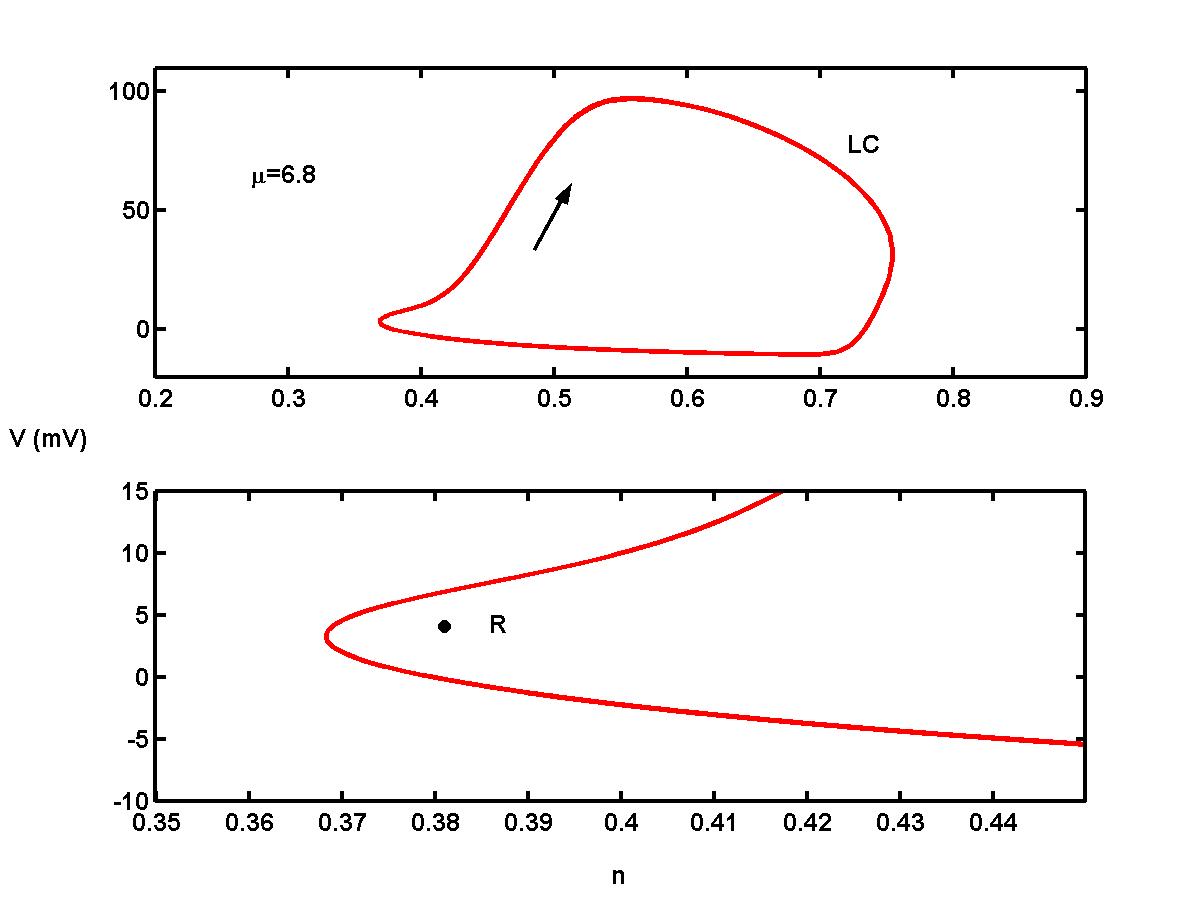}}
\begin{figure}[h]
\caption{In the top part, voltage  is plotted versus potassium activation 
variablefor the deterministic path of repetitive
spikes, depicting the limit cycle for the HH ODE system with
$\mu=6.8 >  \mu_{c_1}$. In the lower part the limit cycle is
magnified in the vicinity of the stable rest point.}
\label{}
\end{figure}

\centerline{\includegraphics[width=4 in]{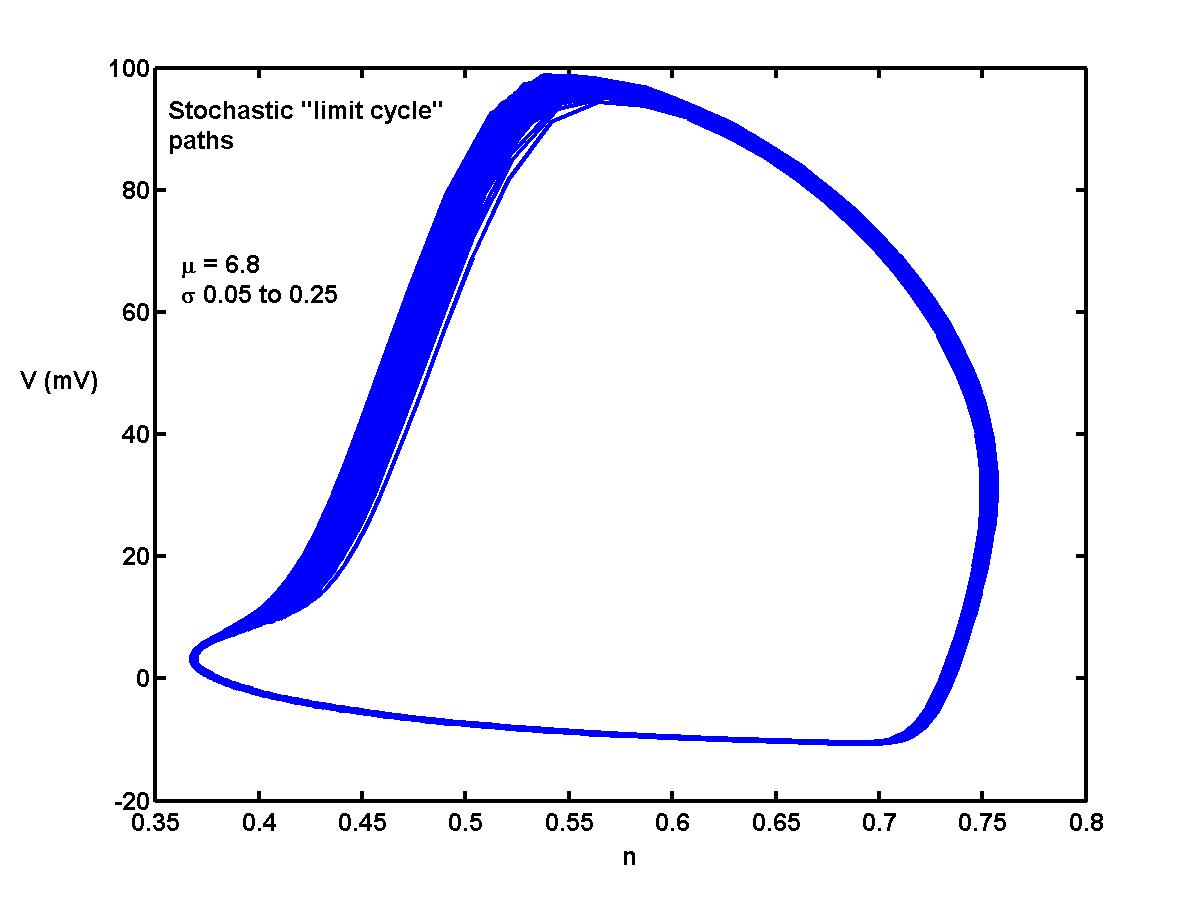}}
\begin{figure}[h]
\caption{The union of many stochastic paths for values of the noise parameter
  $\sigma$ for which paths did not collapse into the stable point in a time interval
of at least 1000 ms.}
\label{}
\end{figure}

Another way to see the limited variability of the times to 
complete a spike orbit (limit cycle) is displayed
 in Figure 7. Here spikes were
observed during repetitive firing during periods in 
which no transitions to the
basin of attraction of the rest point occurred, with noise levels from $\sigma= 0$
 to  $\sigma= 0.4 $. The statistical 
properties of the interspike interval (ISI) are shown in the Figure. 
The most salient feature is that the mean ISI is practically
constant (blue triangles)  as the noise varies, staying in the interval $[17.58, 17.82]$ 
for these values of $\sigma$. Naturally the standard deviation
of the ISI increases (green squares) as $\sigma$ increases, the maximum ISI
increases /red circles) and the minimum ISI decreases (black diamonds).

\centerline{\includegraphics[width=5 in]{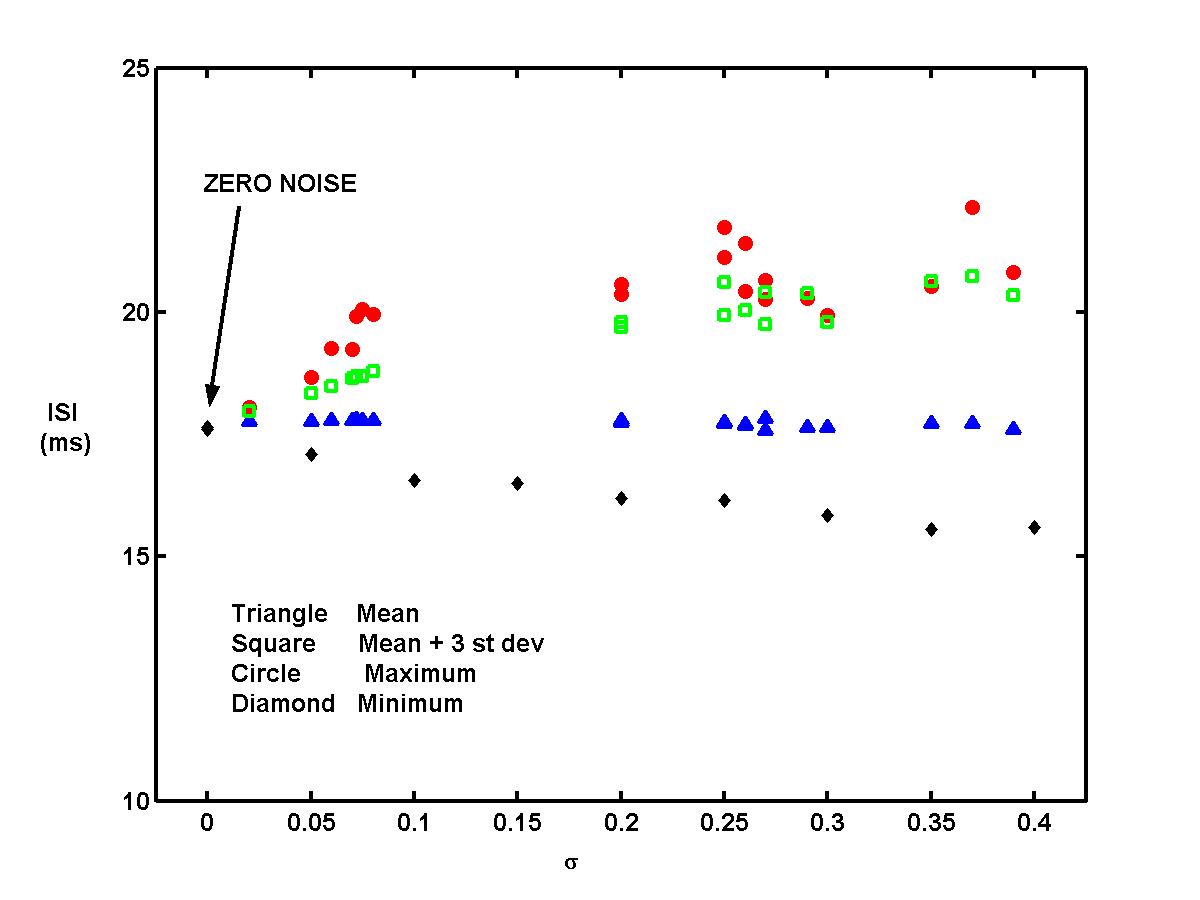}}
\begin{figure}[h]
\caption{Some statistical measures of the ISI during 
repetitive spiking at various noise levels with $\mu=6.8$.
 For $\sigma=0$, there
is no variability. The mean ISI is shown with (blue) triangles,
the (green) squares denote the mean +  3 standard deviations,
 the (red) circles denote the maxima and the (black) diamonds
denote the minima.  Until $\sigma$ is
about 0.07, there are no cessations of spiking 
up to 500000 ms.}
\label{}
\end{figure}

Examples of  distributions of the ISI during repetitive spiking 
at small noise levels are shown in Figure 8. In the top
panel is an ISI histogram for a noise level 
($\sigma=0.07$) at which there
is apparently no cessation of spiking over extremely long
(infinite?)
time periods with 28429 ISIs in 500000 ms.
 The distribution is roughly Gaussian and
has a mean of 17.59 ms and a standard deviation of
0.221 ms. For the second histogram shown, $\sigma=0.085$, 
which is just greater than the critical value at which
spiking may stop in a finite time. The number of spikes is only 4397
and the mean and standard deviation of the ISI are
17.60 ms and 0.276 ms, respectively. The distribution of the
ISI is practically the same in both cases.

\centerline{\includegraphics[width=5 in]{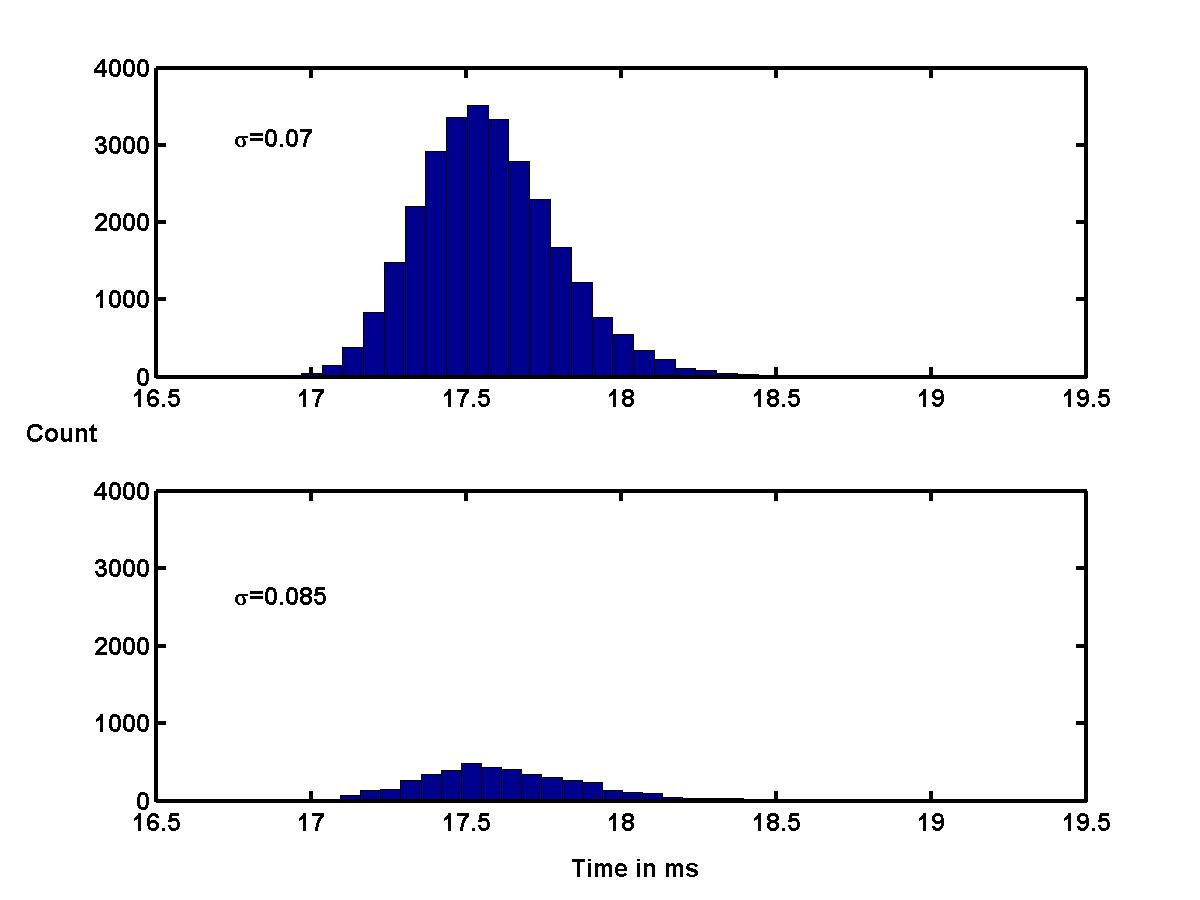}}
\begin{figure}[h]
\caption{Histograms of the ISI during 
repetitive spiking at two small noise levels. Top, $\sigma=0.070$;
bottom, $\sigma=0.085$. For $\sigma=0.07$ there are no cessations of spiking 
up to 500000 ms whereas for $\sigma=0.085$ firing stops after
about 4400 spikes.}
\label{}
\end{figure}

\section{Results on spiking for the HH system and inverse stochastic resonance}
In the following it is assumed that $\mu$ is such that
repetitive spiking does in fact occur. 
In relation to the stochastic paths for the HH system of stochastic
equations we define the following two random variables.

Firstly, the exit time of the process to escape from the basin
of attraction $B_L$ of the limit cycle L to that  $B_R$ of the rest point R is  by
\be T_{L\rightarrow R} (\bs x_0),  \bs x_0 \in B_L  \ee
where $  (\bs x_0) =(V_0, n_0, m_0, h_0)$ is an 
initial point. 
Secondly the exit time for the process to escape from the basin
of attraction  $B_R$ of R to that of the limit cycle is 
\be T_{R \rightarrow L} (\bs x_0),  \bs x_0 \in B_R. \ee
Using the standard theory for diffusion processes, Kolmogorov
second order partial 
differential equations for the moments and distributions of
these quantities as a function of initial
values were described in our previous work (Tuckwell et al., 2009).
However, any attempt to solve these equations analytically or even
by numerical methods for PDEs, apart from
being a formidable task, requires an exact or even 
approximate  knowledge
of the two basins of attraction which is unfortunately not
presently available. In fact the probabilistic nature of $B_L$ and
$B_R$ is not completely understood because it seems that
for small enough noise, at least for some values of $\mu$, 
escape from $B_L$ may be impossible,  not being  observed
in extremely long time periods and similarly, for particular
ranges of $\sigma$, for escape from $B_R$.  

\subsection{Long term trials and data collected}
We are interested in obtaining samples of meaningful
sizes for the above random variables and determining their
properties, mainly as a function of $\sigma$.  Since some of the
exits from one basin of attraction to the other are, for
some values of $\sigma$, very rare events the simulations
of solutions of the full stochastic HH system have to be 
performed over very long time periods. These were chosen
to be 500000 ms or about 8 minutes and 20 seconds with
a timestep of 0.065 ms. 
Results for the 500000 ms interval were obtained as the
union of 100 trials of length 5000 ms, where the final values
of $V,n,m,h$ from any trial were used as the initial values
for the next trial. Then the data on $t, V, n, m, h$ for the 100 trials
were concatenated to give 5 single vectors each with 
 over 7.5 million elements.
The records for $V$ were analyzed to determine the
times of occurrence of spikes and from these the interspike 
intervals were determined. In total, 50 trials of length 500000
were simulated.

\subsection{All spikes}
The numbers of spikes were recorded in each of 50 trials
of length 500000 ms for 40 values of $\sigma$. The mean number
of spikes, denoted by E[NSP],  
 is plotted against $\sigma$ in Figure 9. 
In the top panel all results are included, whereas in the
lower left panel detail is shown for $0 \le \sigma \le 0.1$ and
in the lower right panel, for $0.1 \le \sigma \le 0.5$.

The general form of the plot of E[NSP] versus $\sigma$
is similar to that in Figure 5 of Tuckwell et al. (2009) where 
the total time period was much less at 1000 ms.  Thus, Figure 9 exhibits the
phenomenon of inverse stochastic resonance, which refers
to a firing rate which, as noise level increases, at first
declines to a minimum and then becomes greater.
 However, Figure 9 shows more
detail and reveals 4 distinct regimes marked $R_1$,...,$R_4$.
In region $R_1$, which extends from the deterministic
setting of $\sigma=0$ to very close to
$\sigma=0.07$, the noise is almost without effect and the number of
spikes is always close to that for the zero noise, 28431.   Region $R_2$ 
is characterized by a rapidly falling number of spikes
as  $\sigma$ increases from 0.07 to 0.14. At the latter value
of $\sigma$ the mean number of spikes is 104.8 which is 
about 0.4\% of the maximum value. In region $R_3$,
from $\sigma=0.15$
to $\sigma=0.35$, E[NSP]  stays below 100 with a
broad minimum of about 9.5 spikes (0.03\% of the maximum)
around $\sigma =0.295$ to 0.300.   By $\sigma=0.375$, the beginning
of region $R_4$, the 
mean number of spikes has reached over 120 and then increases
sharply and eventually at a slower rate to reach  25883 around 
$\sigma=2$.

\centerline{\includegraphics[width=5 in]{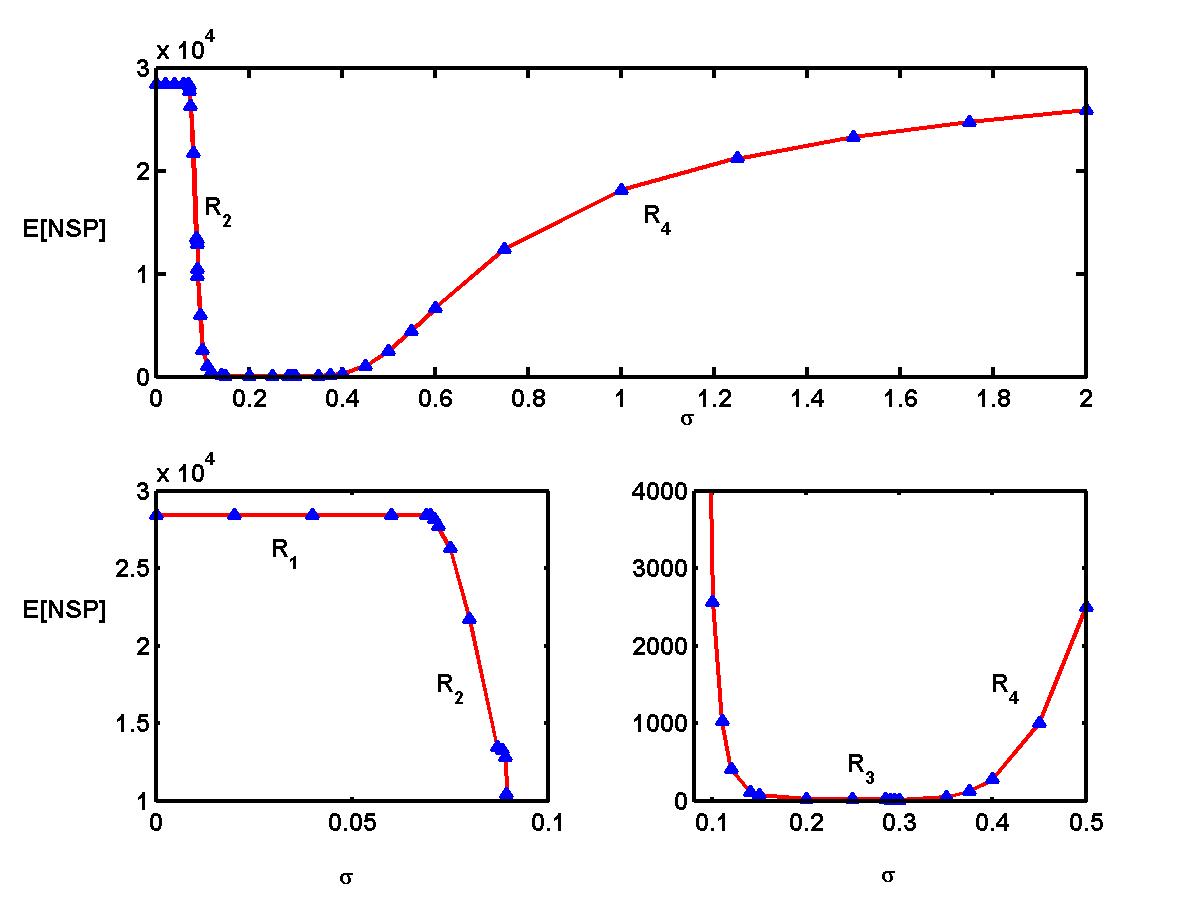}}
\begin{figure}[!h]
\caption{The dependence on the noise parameter
$\sigma$ of the mean of the total number of spikes E[NSP] 
in a time interval of length 500000 ms in the nonlinear HH system.
 Here $\mu=6.8$ with 50 trials at each point.
The values of $\sigma$ are divided into 4 regimes designated $R_1$ to $R_4$.
In the top panel, results are given for all values of $\sigma$. In the lower
left panel the detail of $R_1$ and the start of $R_2$ are shown. In the
lower right panel, the detail of $R_3$ and the start of $R_4$ are shown.  }
\label{}
\end{figure}

\subsection{Underlying scheme for ISR}
The following observations constitute a basis
for ISR which occurs at some values of $\mu$ in the
HH system described by Equations (1)-(4).
We define two critical values  $\sigma_{c_1} $ and $\sigma_{c_2}$ 
of the noise parameter
$\sigma$, with $0 < \sigma_{c_1}   < \sigma_{c_2} < \infty.$ 
$\sigma_{c_1} $ and $\sigma_{c_2}$ depend on $\mu$, and 
are only relevant above the critical value for repetitive firing and
possibly for selected initial values of the process (see Tuckwell et al., 2009).
With reference to the two random variables defined by (26) and (27),
but without any specific values of $ \bs x_0$, we have the following,
as schematized diagrammatically in Figure 10. 

\b  ${ \bs 0 \bs <  \bs \sigma \bs < \bs \sigma_{c_1} }$. \\
Repetitive spiking continues, presumably indefinitely.\\
The probability of a transition from $B_L$ to $B_R$ at any time is zero.\\
The expectation of  $T_{L\rightarrow R} (\bs x_0),  \bs x_0 \in B_L$, is infinite.
\v
\b  $  \bs \sigma_{c_1} \bs < \bs \sigma <   \bs \sigma_{c_2}.$ \\
Repetitive spiking ceases at some time $ t_1<\infty$.\\
The probability of a transition from $B_L$ to $B_R$ is greater than zero and
increases with noise level.\\
The expectation of  $T_{L\rightarrow R} (\bs x_0),  \bs x_0 \in B_L$, is finite and
and tends to decrease with increasing $\sigma$.\\
After a transition from $B_L$ to $B_R$, the probability of the reverse
transition from $B_R$ to $B_L$ is zero.\\
The expectation of  $T_{R\rightarrow L} (\bs x_0),  \bs x_0 \in B_R$, is infinite.\\
After the cessation of repetitive spiking, there is no more spiking.
\v
\b  $ \bs \sigma \bs > \bs \sigma_{c_2} $. \\
After a transition from $B_L$ to $B_R$, the probability of a 
transition from $B_R$ to $B_L$ is greater than zero and tends
to increase as  $\sigma$ incresases.\\
The expectation of $T_{R\rightarrow L} (\bs x_0),  \bs x_0 \in B_R$, is 
finite and tends to decrease as $\sigma$ increases.\\
After a transition back from $B_R$ to $B_L$, the reverse
transition occurs with non-zero pobability, followed again
by a transition from $B_L$ to $B_R$  and so forth, indefinitely.\\
Epochs of spiking and nonspiking alternate with variable durations which 
tend to become shorter as $\sigma$ increases. 

Figure 10 illustrates some of these aspects. The bottom panel
shows part of the actual E[NSP] curve (normalized) from Figure 9 and 
the approximate values of  $ \sigma_{c_1} $ and $ \sigma_{c_2} $
are obtained from this curve. In the top panel, the probabilities
of transitions from  $B_L$ to $B_R$ and $B_R$ to $B_L$ are indicated as
commencing to increase from zero at the two critical values of $\sigma$ and
rising to unity, although this final value depends on the initial value  $\bs x_0 $
which is not fixed throughout. In the middle panel, descriptions of the 
expectations of the exit times from  $B_L$ and $B_R$ are given.

\centerline{\includegraphics[width=5 in]{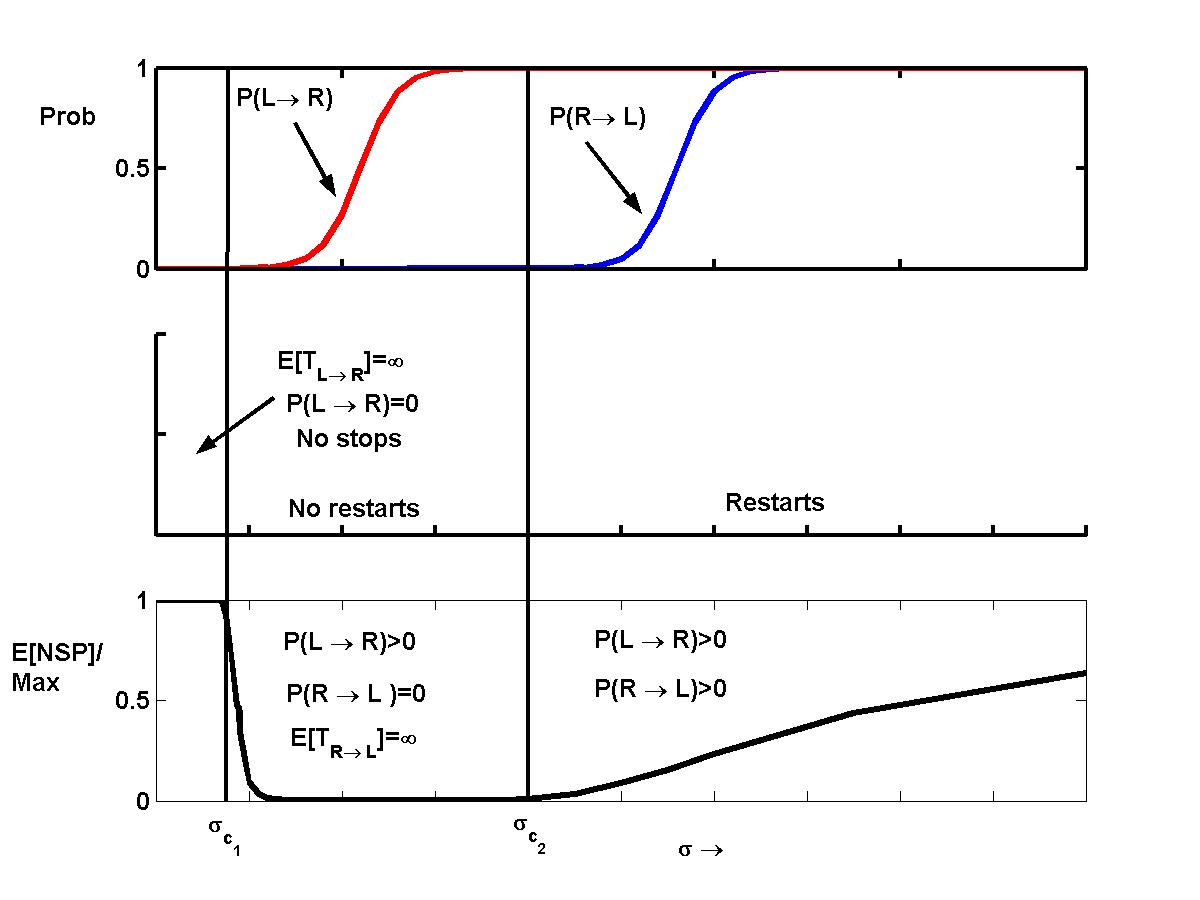}}
\begin{figure}[!h]
\caption{A schematic representation for the transitions underlying
ISR. There are shown two critical values of the noise parameter $\sigma$
and the probabilities of transitions between the basins
of attraction $B_L$ and $B_R$ of the limit cycle and rest point
are sketched in the top panel for the various ranges of values of
$\sigma$. In the bottom panel is shown the normalized expected number
of spikes obtained by simulation over 500000 ms for $\mu=6.8$, a value
of $\mu$ at which ISR had been found to be pronounced.}
\label{}
\end{figure}

The scheme given in Figure 10 is in accordance with the
results for sample paths of $V$ for simulated spike trains shown in Figure 11. 
Here the time periods
are all 500000 ms but in the top two records time is only shown to
2000 ms.  The very small noise case gives apparently
indefinite repetitive spiking. Somewhat larger noise leads to
an initial isolated  burst
followed by silence, possibly for an infinite time. 
Increasing the noise further leads to occasional bursting
and eventually the frequent bursting case occurs.

\centerline{\includegraphics[width=5 in]{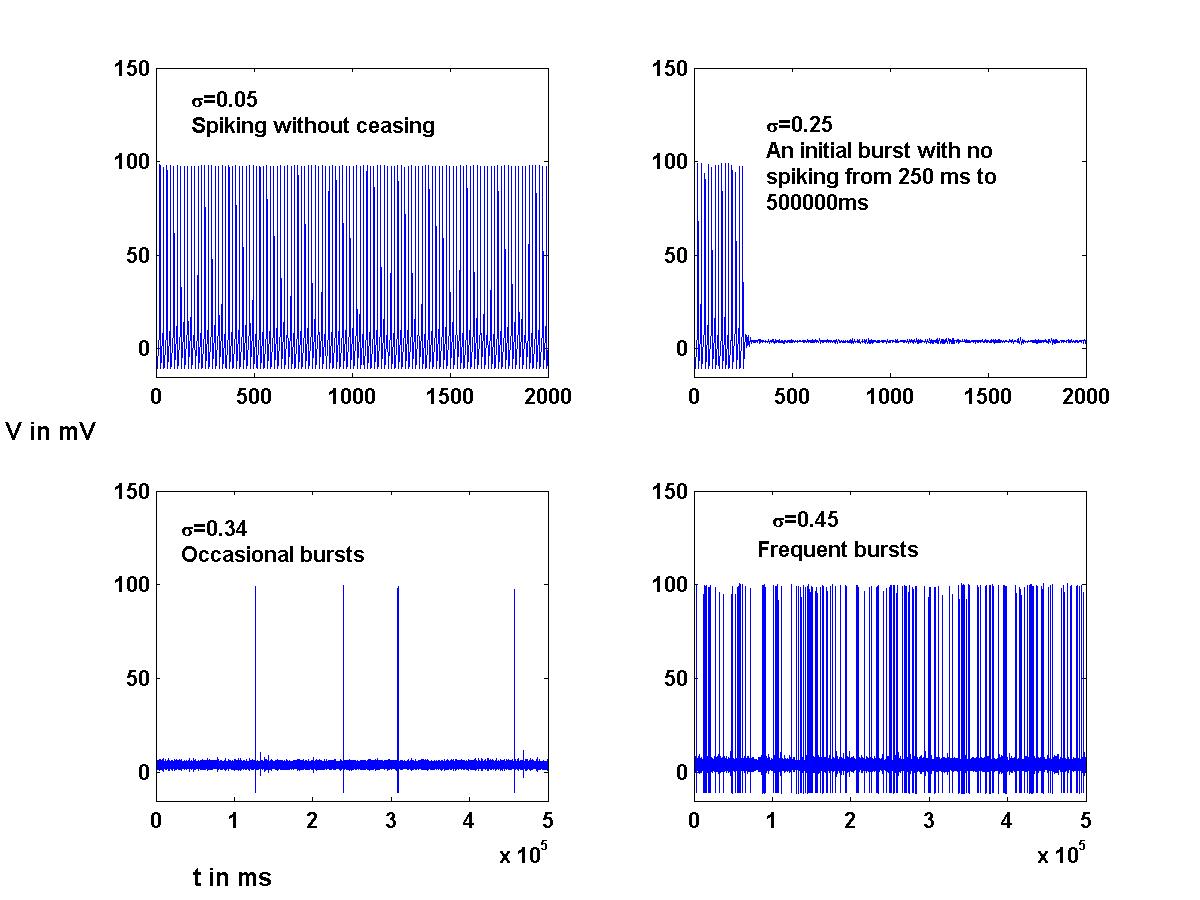}}
\begin{figure}[!h]
\caption{Illustrating the 4 basic patterns of spiking activity.
Top left: $\sigma=0.05$.  Incessant spiking, at least to 500000 ms.
Top right: $\sigma=0.25$. One initial burst with no further spikes, at least to 500000 ms.
Bottom left: $\sigma=0.34$. Occasional bursts with separations of order 100000 ms.
Bottom right:  $\sigma=0.45$. Frequent bursts.  }
\label{}
\end{figure}

\subsection{Some statistics of $T_{L\rightarrow R}$ and $T_{R\rightarrow L}$}
In this subsection results on certain statistical aspects 
of the underlying random variables
 $T_{L\rightarrow R}$ and $T_{R\rightarrow L}$, defined by (26) and (27),
 obtained from the long-term simulation
of the HH system (1)-(4) are given. 

\subsubsection{Escape time from $B_L$}
During repetitive spiking the trajectories of the process
do not deviate much from the deterministic limit cycle as illustrated
in Figure 6.  It is assumed that there is a well-defined but unkown
set in $(V,n,m,h)$-space, denoted by $B_L$, containing the
deterministic repetitive spiking trajectory, which is called the basin
of attraction of the limit cycle. This implies that if the process
started in $B_L$, then, with no noise, the path would approach
the limit cycle. With noise, trajectories may escape from $B_L$ 
whereupon they enter the basin of attraction $B_R$ of the stable
equilibrium point. The nature of $B_L$ is unknown 
and it is not clear that it is a regular open set because
if it was it is likely that escape from it would occur eventually
with probability one no matter how small the noise level.
Thus, even  though it seems that for  $\sigma < \sigma_{c_1}$, 
$P(L\rightarrow R)=0$ and  E[$T_{L\rightarrow R} =\infty$],
it may be the case that $P(L\rightarrow R)$ is so small for
small enough $\sigma$ that the event of this escape is unlikely
to be observed in the course of feasible simulations. 
It is a remaining mathematical challenge to ascertain the
veracity of these remarks. 

Notwithstanding these uncertainties, the mean value of 
 E[$T_{L\rightarrow R}$] of the exit time from $B_L$ 
was estimated from sample paths and the results are shown for
$\sigma \ge 0.2$ in Figure 12. For these noise levels the mean
exit time declines rapidly from very large values to a minimum
of about 57 ms at $\sigma=1.25$.   Thereafter,
 E[$T_{L\rightarrow R}$]  seems to increase slightly to  about
72 ms at $\sigma=2$.

\centerline{\includegraphics[width=5 in]{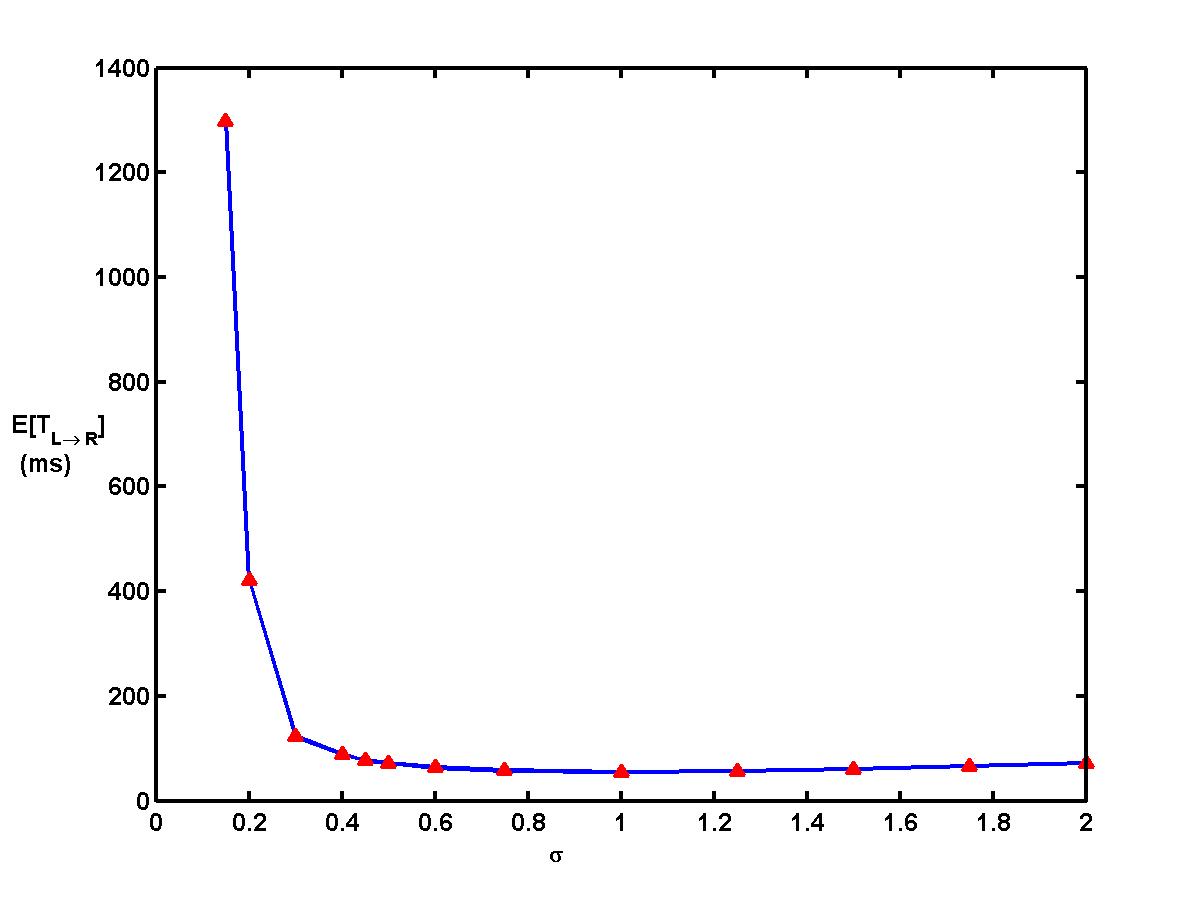}}
\begin{figure}[h]
\caption{The dependence on the noise parameter
$\sigma$ of the expectation E[$T_{L\rightarrow R}$] of the random variable which
is the time of exit of the process from the basin of attraction of the limit cycle
to that of the
stable rest point. 50 trials at each point with $\mu=6.8$. 
For values of $\sigma $ less than a critical value $\sigma_{c_1} \approx 0.07$
 the value of this expectation is apparently infinite. Extremely large
values which occurred for $ \sigma_{c_1}  < \sigma <0.2$ are not shown.}
\label{}
\end{figure} 
\subsubsection{Escape time from $B_R$}
The basin of attraction of the stable rest point is also unknown
exactly.  Figures 13 and 14 show estimates of the mean and  
standard deviation of $T_{R\rightarrow L}$ over various ranges
of values of $\sigma$. Again, it is not known with certainty, 
but it appears that the probability of escape from $B_R$ to
$B_L$ is zero (or extremely close to zero) until $\sigma \ge \sigma_{c_2}$.
In Figure 13, the mean is shown dropping from values of order
300000 ms at $\sigma=0.25$ to eventually reach values about 30 ms at
$\sigma=2$.  Figure 14 shows corresponding results for the standard
deviation, but only for $\sigma \ge 0.35$ because sample sizes
were too small for lesser values of the noise parameter. 
In the lower panel of Figure 14 is shown the dependence
of the coefficient of variation of  $T_{R\rightarrow L}$ on noise level.
It can be seen that there is an initial increase in this quantity until
a maximum is attained at about $\sigma=0.5$, whereupon it
declines monotonically.

\centerline{\includegraphics[width=5 in]{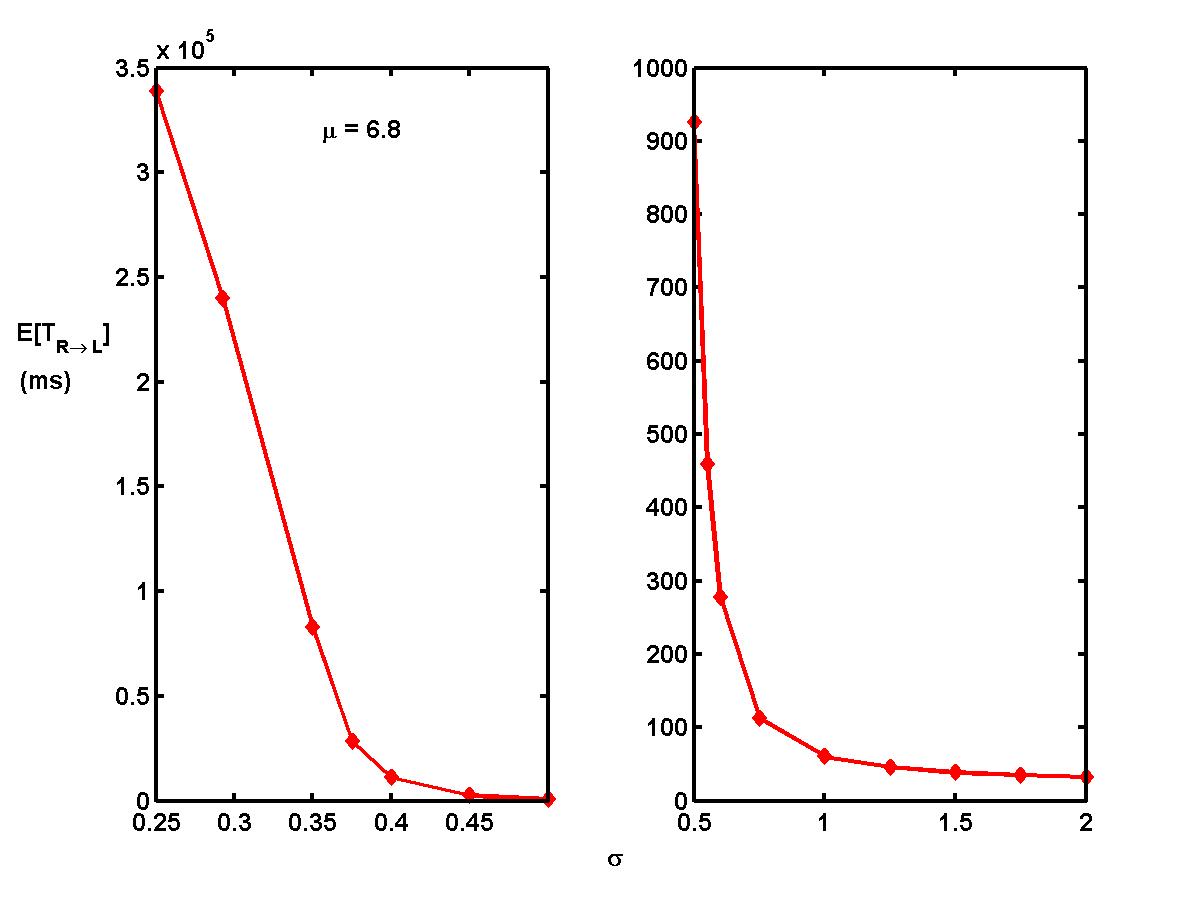}}
\begin{figure}[h]
\caption{The dependence on the noise parameter
$\sigma$ of the expectation E[$T_{R\rightarrow L}$] of the random variable which
is the time of exit of the process from the basin of attraction of the
stable rest point to that of the limit cycle. Here $\mu=6.8$. 50 trials at each point.
For values of $\sigma $ less than some value just less than 0.25
 the value of this expectation is apparently infinite. }
\label{}
\end{figure}

\centerline{\includegraphics[width=5 in]{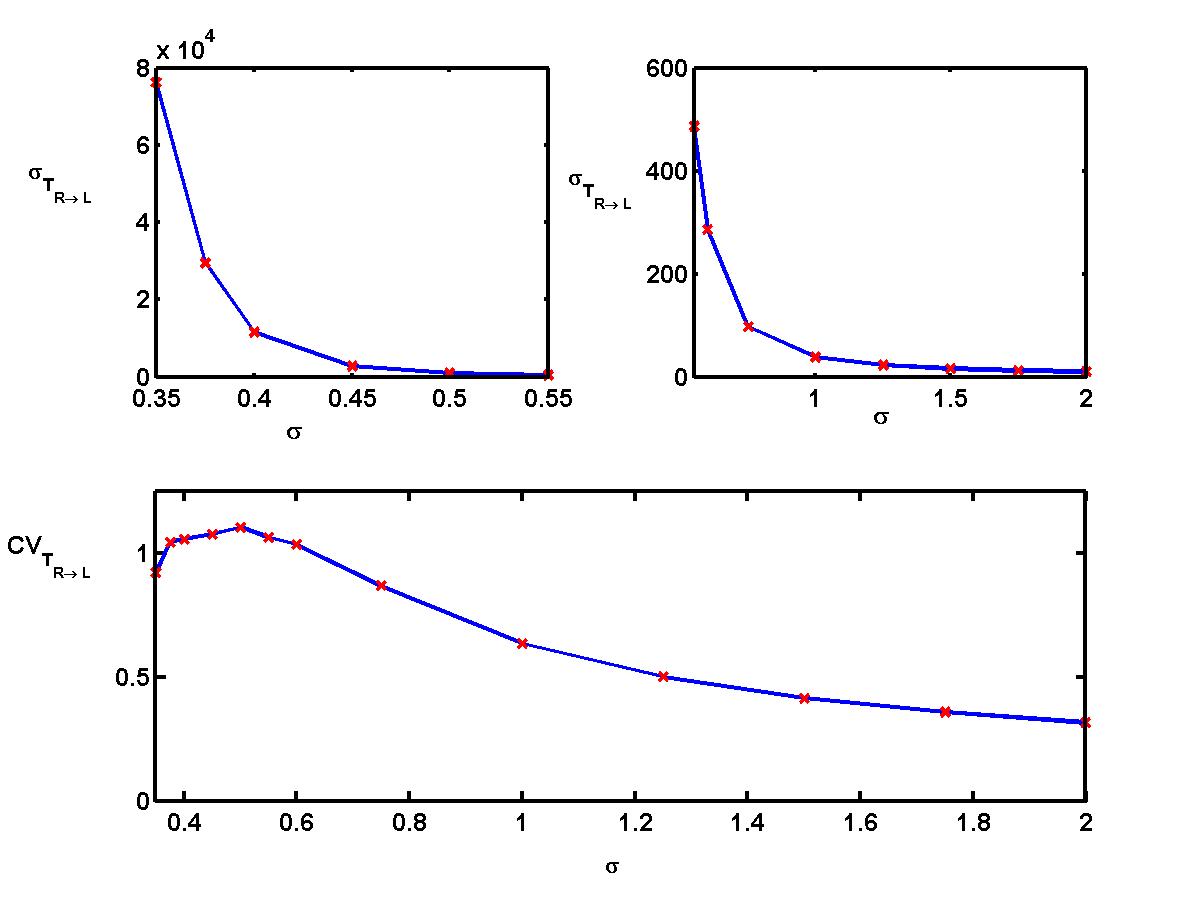}}
\begin{figure}[h]
\caption{{\it Top two panels}. The dependence on the noise parameter
$\sigma$ of the standard deviation $\sigma_{T_{R\rightarrow L}}$,of the random variable which
is the time of exit of the process from the basin of attraction of the
stable rest point to that of the limit cycle. Here $\mu=6.8$. 50 trials at each point.
{\it Bottom panel}. The coefficient of variation of the random variable $T_{R\rightarrow L}$
as a function of $\sigma$.}
\label{}
\end{figure}

\subsubsection{Distributions}
There are three  main random variables of interest in relation to the empirical spike
trains  obtained in this study. These are the previously defined exit times 
$T_{L\rightarrow R}$ and $T_{R\rightarrow L}$ and in addition
the general ISI for the whole train. Samples for the latter random variable
can be viewed as the union of those of the previous two. 
Examples of histograms of these random variables are shown in Figure 15.
In all cases the results for 50 trials of duration 500000 ms are combined. 
In Figure 15A is given a histogram for all ISIs for $\sigma=0.2$. This occurs in
region $R_3$ of Figure 9, where there are very few spikes, all with short
ISIs as there are no returns to $B_L$ fron $B_R$. The histogram is of
the same nature as those in Figure 8. See also Figure 7. 

In Figure 15C  the raw histogram is shown for all ISIs with $\sigma=0.350$. 
There is a preponderance of small intervals and then an exponential-type
distribution of longer ($>21.5$ ms) intervals as depicted in Figure 15E.
The tail of the exponential distribution is very long and extends out
to about 500000 ms which is the limit in these simulations.  The distribution
depicted in Figure 15E is close to the actual distribution of $T_{R\rightarrow L}$ 
for this value of $\sigma$. Similar remarks apply to the pairs of Figures 15B and
15D ($\sigma=0.375$) and 15F and 15G ($\sigma=0.5$). 
In the final plot of Figure 15H, the number of spikes has become
enormous and the majority of ISIs are less than 50 ms as expected
for $\sigma=1$ which is in region $R_4$ of Figure 9.
In the histograms of Figure 15 C, B, F and H, there are 
small bin counts out to very large times and these are
not visible compared to the large bin count at short intervals. 
However, they are visible at relatively small values in the truncated histograms. 

\centerline{\includegraphics[width=5.5 in]{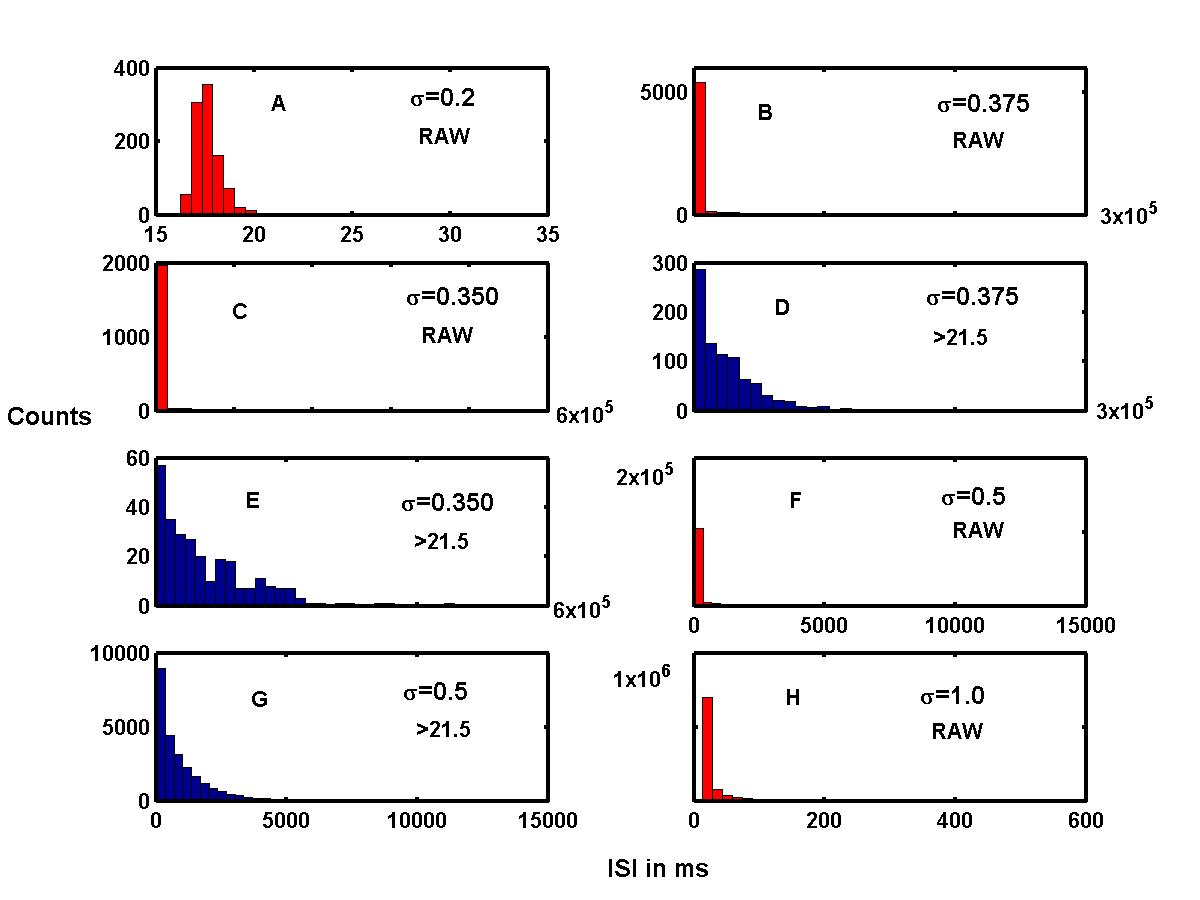}}
\begin{figure}[h]
\caption{Examples of distributions (histograms) of ISIs and exit times 
from $B_R$ based on 50 trials (data pooled)
of length 500000 ms. Red histograms, raw data. Blue histograms, ISIs truncated at 21.5 ms.
A.  Raw data for  $\sigma=0.2$. B. Raw data for $\sigma=0.375$.
C. Raw data for  $\sigma=0.350$.  D. Truncated data for $\sigma=0.375$. 
E. Truncated data for $\sigma=0.350$. F.  Raw data for  $\sigma=0.5$.
G. Truncated data for $\sigma=0.5$. H.  Raw data for  $\sigma=1.0$.}
\label{}
\end{figure}

To complete the picture for the distributions of the three 
key random variables,  Figure 16 shows histograms of mean numbers of
spikes per burst for two values, 0.4 and 1.0, of $\sigma$.  If these are multiplied 
by the mean ISI within bursts, which is about 17.6 ms, the mean times
spent in the basin of attraction of the limit cycle ($B_L$) and hence of 
 $T_{L\rightarrow R}$ are estimated. For smaller $\sigma$ some very large 
means were obtained (not shown). Noticeable in Figure 16 is the
smaller magnitude and 
small variability when $\sigma=1.0$ compared to $\sigma=0.4$.
The distributions in both cases shown are approximately Gaussian
which may be compared with the exponential-types shown in 
Figures 15 D, E and G.


 \centerline{\includegraphics[width=5.0 in]{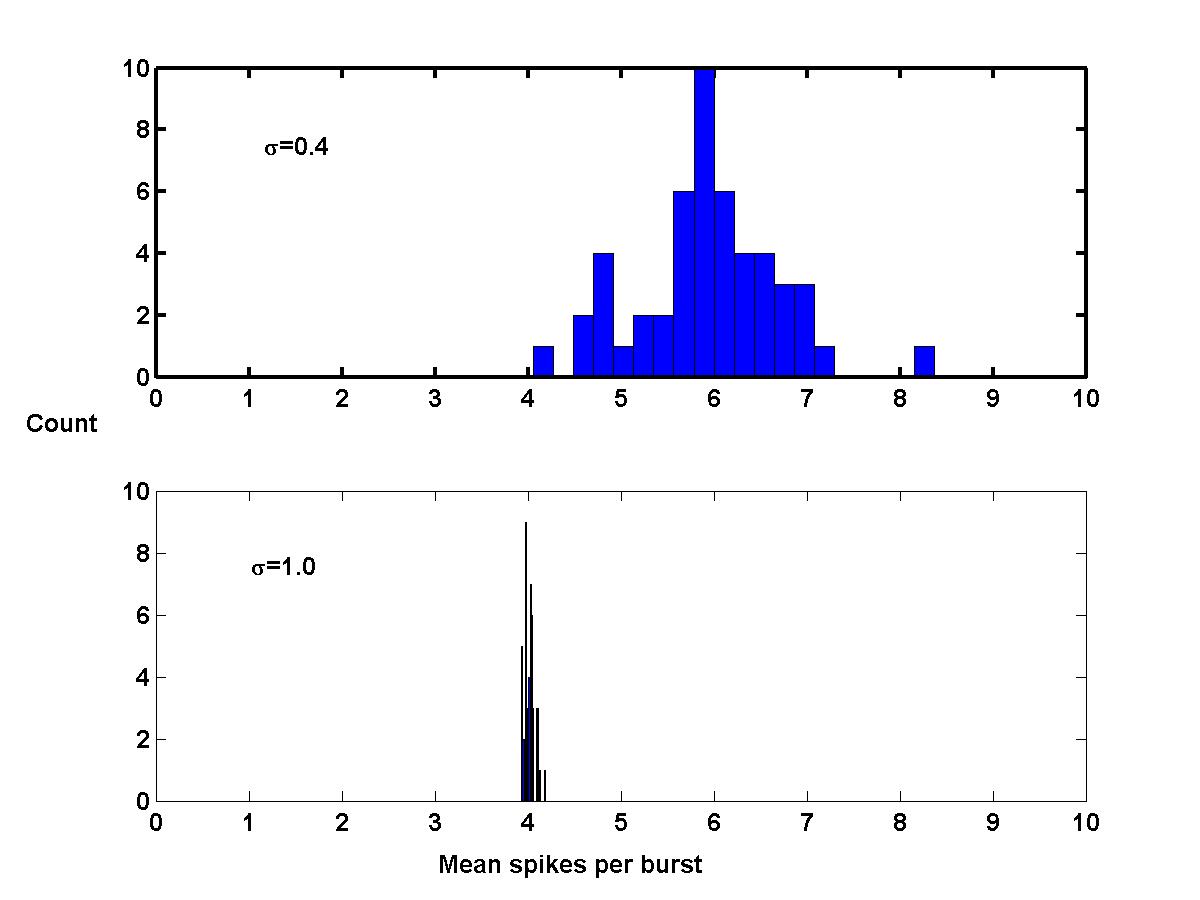}}
\begin{figure}[h]
\caption{Histograms of mean numbers of  spikes per burst,
which indicate magnitudes of exit times from $B_L$ to $B_R$.
based on 50 trials
of length 500000.}
\label{}
\end{figure}

\section{Discussion}
The inhibitory effects of noise on repetitive spiking 
in squid axon, on which the HH system is based,
were well documented experimentally by Paydarfar et al.  
\cite{PAY}.  The first theoretical evidence of ISR in the full HH sytem
was obtained
for $\mu=5$ in the non-repetitive spiking mode \cite{TUC5} (Figure 2),
as it was found that there was a maximum in the mean ISI
at small values of $\sigma$. Subsequently ISR has been 
demonstrated for repetitive spiking in both the
HH system of ODEs and PDEs  \cite{TUC1, TUC2, TUC3, GUO}.
In addition to using the standard initial conditions and additive noise, 
in \cite{TUC1} the phenomenon was confirmed with respect
to random initial conditions and conductance-based input.
Some theory of  ISI was also outlined \cite{TUC1} in terms of bifurcations in the
HH sytem 
\cite{HAS,FUK,JOS}  and 
of the exit times \cite{TBK2} from 
the basins
of attraction of a stable equilibrium point and  a limit cycle.
However the simulations were over relatively short time
periods and the statistical details of the exit times were not
explored in detail.

In this article we have first explored the properties of the stable
equilibrium point in depth. Linearizing the stochastic  HH system about that point
gave an approximate sytem of stochastic differential equations
whose oscillatory solutions were found to mimic those of 
the full system in nonspiking periods. The oscillations
are evidently an integral part of firing in the HH system
with the considered parameters, because spikes did tend to emerge
at the maxima. The distribution of ISIs with trajectories
not departing greatly from the limit cycle 
for very small noise was estimated and is, as perhaps
 expected, Gaussian-like.

Long-term simulations, to 500000 ms were performed with $\mu=6.8$ 
for
many values of the noise parameter $\sigma$. Four ranges of 
values of $\sigma$ were distinguished, based on mean
total numbers of spikes. These regions were denoted $R_1,...,R_4$
in Figure 9. Two critical values of $\sigma$ were also apparent, denoted
by  $ 0 < \sigma_{c_1} < \sigma_{c_2} <\infty$.
A detailed discussion of the mechanisms underlying ISR in terms 
of these  critical values was given  in Section 3.3. 
It is not certain whether escape from the basin of attraction
of the limit cycle can ever occur when $\sigma < \sigma_{c_1} $
or whether escape from the basin of attraction from the
basin of attraction of the stable equilibrium point can ever occur 
when $\sigma_{c_1}  < \sigma <  \sigma_{c_2}$, but these events
were never observed in 50 trials of length 500000 ms with several
values of $\sigma$.

 \end{document}